\documentclass[12pt]{iopart}

\usepackage[svgnames]{xcolor} % Specify colors by their 'svgnames', for a full list of all colors available see here: http://www.latextemplates.com/svgnames-colors
\usepackage{graphicx,epsfig}
\usepackage{times}
\usepackage{xcolor}
\usepackage[normalem]{ulem}
\usepackage{lipsum}

\usepackage{float}

\definecolor{albicocca}{rgb}{0.98, 0.7, 0.2}
\definecolor{internationalorange}{rgb}{1.0, 0.31, 0.0}
\definecolor{giocolor}{RGB}{0, 150, 100}

\begin{document}

\title{Pathways to explosive transitions in interacting contagion dynamics}

\author{Santiago Lamata-Otín}
\address{Department of Condensed Matter Physics, University of Zaragoza, 50009 Zaragoza, Spain}
\address{GOTHAM lab, Institute of Biocomputation and Physics of
Complex Systems (BIFI), University of Zaragoza, 50018 Zaragoza, Spain}

\author{Jes\'us G\'omez-Garde\~nes}
\address{Department of Condensed Matter Physics, University of Zaragoza, 50009 Zaragoza, Spain}
\address{GOTHAM lab, Institute of Biocomputation and Physics of
Complex Systems (BIFI), University of Zaragoza, 50018 Zaragoza, Spain}
%\affiliation{Center for Computational Social Science, University of Kobe, 657-8501 Kobe, Japan}

\author{David Soriano-Pa\~nos}
\address{Instituto Gulbenkian de Ciência, 2780-156 Oeiras, Portugal}
\address{GOTHAM lab, Institute of Biocomputation and Physics of
Complex Systems (BIFI), University of Zaragoza, 50018 Zaragoza, Spain}
\ead{sorianopanos@gmail.com}
\vspace{10pt}
\begin{indented}
\item[]November 2023
\end{indented}

\begin{abstract}
Yet often neglected, dynamical interdependencies between concomitant contagion processes can alter their intrinsic equilibria and bifurcations. A particular case of interest for disease control is the emergence of explosive transitions in epidemic dynamics coming from their interactions with other simultaneous processes. To address this problem, here we propose a framework coupling a standard epidemic dynamics with another contagion process, presenting a tunable parameter shaping the nature of its transitions. Our model retrieves well-known results in the literature, such as the existence of first-order transitions arising from the mutual cooperation of epidemics or the onset of explosive transitions when social contagions unidirectionally drive epidemics. We also reveal that negative feedback loops between simultaneous dynamical processes might suppress explosive phenomena, thus increasing systems robustness against external perturbations. Our results render a general perspective towards finding different pathways to explosive phenomena from the interaction of contagion processes.
\end{abstract}

%\maketitle

\section{Introduction}

Understanding complex systems requires shifting from reductionist theories, devoted to the study of their components in isolation, towards formalisms addressing the crucial role of interactions in the emergence of collective behaviors~\cite{anderson1972more,bianconi2023complex}. The study of the flocking behavior observed in groups of birds \cite{toner1998flocks,bialek2012statistical} or the synchronous firing of a collection of interacting neurons \cite{orlandi2013noise,faci2019impact} are some examples illustrating the successful application of complex systems science to characterize a myriad of collective phenomena occurring across different scales in nature. Nonetheless, analogously to the failure of reductionist theories in understanding complex systems, focusing on single isolated dynamics might fail in capturing the phenomenology resulting from their interactions with other concomitant dynamical processes. The latter issue calls for the development of new integrating frameworks, modeling the possible interdependencies \cite{danziger2019dynamic,castioni2021critical} connecting dynamical processes unfolding across similar time scales and their outcome in the observed collective states.

Undoubtedly, one of the most paradigmatic illustrations of the impact of such interdependencies is the comorbidity of infectious diseases. Some examples are the increased vulnerability to several diseases following HIV infection \cite{HIV,pawlowski2012tuberculosis} or the surge in fatalities in patients with pneumonia during the Spanish Flu outbreak \cite{influenza}. Alternatively, the presence of one pathogen in the population might hamper the propagation of another one. Such competition is usually observed in the epidemic trajectories of viruses with multiple strains circulating simultaneously, such as influenza \cite{influ} and DENV \cite{DENV} or more recently in the complex landscape depicted by the different SARS-CoV-2 variants \cite{barreiro2022modelling,chen2022global}. Outside the domain of interacting epidemics, another clear example of the relevance of interdependencies among simultaneous dynamics is the influence of social behavior on epidemic spreading \cite{Funk2010,Wang2016}. For instance, the existence of mutual feedback between the individual adoption of preemption measures and the spread of a pathogen \cite{khanjanianpak2022emergence,manrubia2022individual} provides a natural mechanism for the emergence of oscillations in the epidemic curves~\cite{steinegger2022behavioural,tang2022controlling}, even in the absence of seasonal effects shaping the intrinsic transmissibility of the virus.

The most conventional approach to model interacting epidemics relies on integrating multiple dynamical processes within a single compartmental model \cite{PhysRevX.4.041005,coupledsis}. Following this approach, numerous studies have explored the effects of competition \cite{PhysRevE.84.036106,pinotti2020interplay} and cooperation \cite{avalanche,PhysRevE.96.022301} in the intertwined dynamics, modeled respectively as a reduction or an increase of the vulnerability of infected population to the other circulating pathogen. One of the most significant findings when addressing the mutual cooperation between epidemic spreading processes is the change in the nature of the transition between the disease-free and epidemic states. Specifically, the second-order transitions characterizing isolated epidemic dynamics turn into first-order ones, yielding the bi-directional enhancement of epidemic processes as a natural pathway towards the emergence of explosive transitions \cite{chen,review_exptran}.

Unlike epidemic processes, social dynamics display first-order transitions in isolation as a result of the complex contagion mechanisms responsible for their diffusion \cite{granovetter1978threshold,simplagion}. This complex contagion reflects that transmission in social dynamics not only depends on the pairwise interaction between the transmitter and the receptor of the information but also on the context of the latter \cite{berger2016contagious,socialimpact}. Mathematically, complex contagions are modeled through synergistic transmission rates \cite{Exp_Tran}, or following higher-order approaches that explicitly incorporate the influence of group structures in the transmission process~\cite{report_Ho,otra_review,bick2022higherorder}. These features prompt the emergence of explosive transitions in epidemic dynamics when being coupled with social dynamics, even in the absence of the aforementioned bi-directional enhancement. Indeed, a recent study by Lucas et al. \cite{unidirectional} shows that higher-order social dynamics, driving but not receiving any feedback from epidemics, induce an explosive transition between the disease-free and the epidemic states.

Regardless of their microscopic origin, the presence of explosive transitions compromises systems' stability, as small perturbations in the system features or negligible stochastic fluctuations can lead to vastly different dynamical outcomes. Hence, the characterization of explosive phenomena has been an important area of research within complex systems over the past few decades \cite{doi:10.1080/00018732.2019.1650450}. Following this line, in this work we are interested in unveiling the possible pathways for the emergence of explosive transitions resulting from coupling two simultaneous spreading processes. For this purpose, we propose a general framework coupling a standard epidemic dynamics with another spreading process whose nature depends on a tunable parameter. Such parameter allows us to explore different contagion processes, ranging from epidemics to social dynamics with strong group effects. Our model aims at addressing (\textit{i}) how explosiveness naturally arises from the interaction between dynamics and (\textit{ii}) what kind of interactions are driving such emergence as a function of the nature of the processes involved.

Our paper is organized as follows: In Section II, we present our general framework to explore pathways to explosive transitions in intertwined spreading processes. In Section III, we delve into the classical interaction between two epidemic dynamics. We confirm the emergence of explosive transitions arising from their mutual cooperation and the second-order extinction of weakest variants induced by their competition with strongest ones. In Section IV, we study the coupling of social dynamics with epidemic processes. First, we consider the unidirectional influence of social dynamics on epidemics. There, we recover the induced explosive transitions in the latter one and characterize analytically the bifurcation points. Secondly, we explore more complex scenarios involving the interactions between explosive and non-explosive processes. We show that explosive transitions are suppressed when coupling both dynamics through a negative feedback loop, highlighting the relevance of these structures for systems' stability. Finally, we report that a mutually competitive scenario involving complex contagions also exhibits discontinuous transitions.

\section{Interacting contagion processes}

\begin{figure*}[t!]
\centering\includegraphics[width=0.9\linewidth]{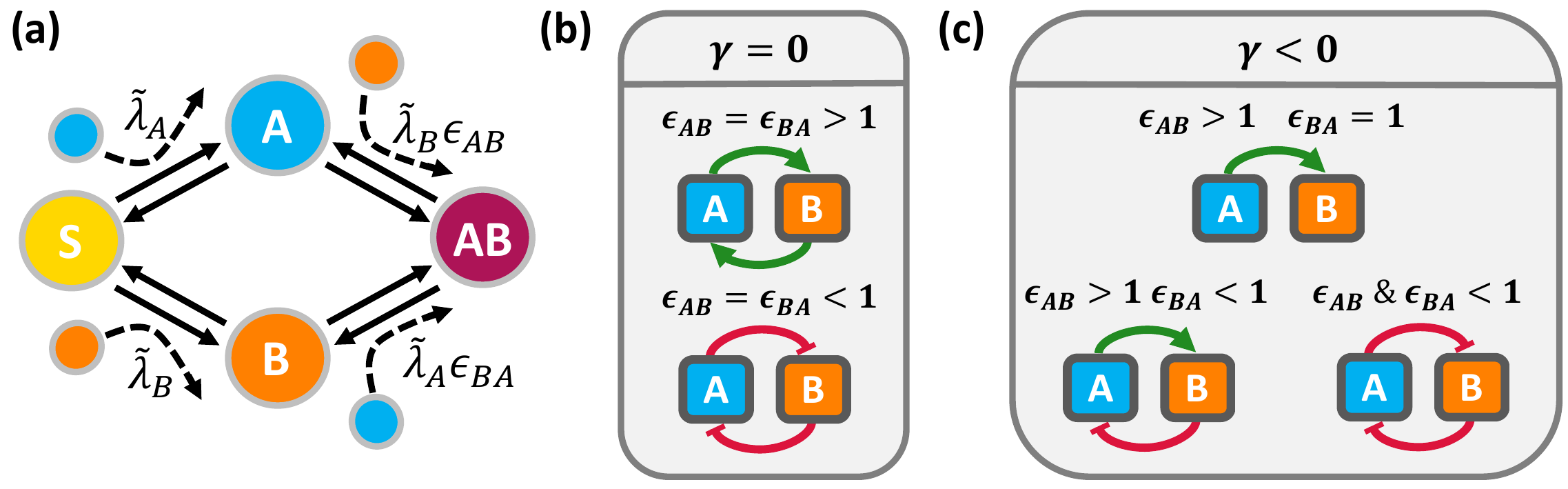}
\caption{\textbf{Compartmental model and interplay between interacting spreading dynamics.} (a) Interacting diseases compartmental model is described. The model has four compartments: susceptible ($S$), infected by pathogen $A$ ($A$), infected by pathogen $B$ ($B$) and infected by both pathogens ($AB$). Arrows indicate the possible transitions between different states. (b) Scenarios explored when coupling two epidemic processes ($\gamma=0$). Top row represent mutually cooperating diseases, i.e. $\epsilon_{AB}=\epsilon_{BA}>1$, whereas bottom row sketches a competitive interaction, i.e. $\epsilon_{AB}=\epsilon_{BA}<1$. (c) Three interactions schemes for a social dynamics $A$ ($\gamma\neq 0$) interacting with another epidemic dynamics $B$: unidirectional enhancement with no feedback, i.e $\epsilon_{AB}>1$ and $\epsilon_{BA}=1$, negative feedback loops, i.e $\epsilon_{AB}>1$ and $\epsilon_{BA}<1$, and mutual competition, i.e $\epsilon_{AB}<1$ and $\epsilon_{BA}<1$.}
\label{fig:panel1} 
\end{figure*}

In this section, we introduce our framework to capture the spread of interacting contagion processes. We assume that there exist two spreading dynamics, denoted by dynamics $A$ and $B$ respectively, which can be modeled as standard Susceptible-Infected-Susceptible (SIS) dynamics in isolation. The SIS model divides the population into two distinct categories: Susceptible ($S$), i.e. those vulnerable to the pathogen (spreading unit), and Infected ($I$), i.e. those who were infected and can transmit the pathogen to other susceptible counterparts. To couple the two simultaneous contagion processes $A$ and $B$, we extend the usual SIS model to accommodate four possible states: Susceptible $S$, infected by just the first pathogen $A$ or the second one $B$ and finally those who are infected by both, denoted by $AB$ \cite{coupledsis,unidirectional}.  The flow diagram of the compartmental model is depicted in Fig.~\ref{fig:panel1}a. First, we assume that susceptible individuals $S$ get infected with disease $A$ or $B$ at a rate $\tilde{\lambda}_A$ or $\tilde{\lambda}_B$ respectively. For the sake of simplicity, we assume the same recovery rate for both diseases, denoted by $\mu$. To account for the interaction between spreading processes, we assume that one individual infected with $A$ ($B$) contracts the pathogen $B$ ($A$) at a rate $\tilde{\lambda}_B \epsilon_{AB}$ ($\tilde{\lambda}_A \epsilon_{BA}$). Therefore, $\epsilon_{ij}=1$ encodes the spread of non-interacting processes, as the infection rates become independent on the dynamical states. Conversely, if $\epsilon_{ij}>1$, the presence of $i$ it enhances $j$ infection rate (cooperation or excitation), and in case $\epsilon_{ij}<1$, $i$ suppresses $j$ infection rate (competition or inhibition).

We consider a mean-field scenario and assume a closed population of $N$ individuals, which allows us to describe the proposed dynamics with a set of three equations. These equations consider the fraction of the population in each infectious compartment, denoted by $\rho_A = A/N$, $\rho_B = B/N$ and $\rho_{AB} = AB/N$ respectively. Without loss of generality, we assume $\mu=1$, obtaining the following set of equations describing the evolution of the system:
\begin{eqnarray}
\dot \rho_A & = & -\rho_A+\tilde{\lambda}_A \rho_S (\rho_A+\rho_{AB})+\rho_{AB}\nonumber \\
             &  & -\epsilon_{AB}\tilde{\lambda}_B \rho_A(\rho_B+\rho_{AB})\;,
\label{eq:a}\\
\dot \rho_B & = & -\rho_B+\tilde{\lambda}_B \rho_S (\rho_B+\rho_{AB})+\rho_{AB}\nonumber \\
             &  & -\epsilon_{BA}\tilde{\lambda}_A \rho_B(\rho_A+\rho_{AB})\;,
\label{eq:b}\\
\dot \rho_{AB} & = & -2\rho_{AB}+\epsilon_{AB}\tilde{\lambda}_B \rho_A(\rho_B+\rho_{AB})\nonumber \\
             &  & +\epsilon_{BA}\tilde{\lambda}_A \rho_B(\rho_A+\rho_{AB})\;,
\label{eq:ab}
\end{eqnarray}
where $\rho_S= 1-\rho_A - \rho_B - \rho_{AB}$ represents the fraction of population in the susceptible state.

The individual contagion rates, $\tilde{\lambda}_A$ and $\tilde{\lambda}_B$ respectively, are set according to the nature of each contagion process. For dynamics $B$, we assume an epidemic process with a constant contagion rate per contact, implying $\tilde{\lambda}_B = \lambda_B$. In isolation, dynamics $B$ undergoes a second-order transition at $\lambda_B=1$ from the spreaders-free state to an endemic equilibrium. Conversely, we introduce a synergistic contagion rate for the spreading process $A$, which reads \cite{Exp_Tran}:
\begin{eqnarray}
    \tilde{\lambda}_A =\lambda_Ae^{\gamma(1-\rho_{A_{T}})}\; ,
    \label{eq:lambdaA}
\end{eqnarray}
with $\gamma\leq 0$. The former expression captures the impact of group pressure on social contagions, as the probability that a susceptible individual starts spreading the idea increases with the fraction of spreaders in the population $\rho_{A_{T}}$. The parameter $\gamma$ allows tuning the nature of the spreading process, ranging from epidemic dynamics ($\gamma=0$) to social contagions with strong group effects as $\gamma$ is decreased. This synergistic contagion rate changes the nature of the dynamics, as there exists a critical value $\gamma_c$~\cite{Exp_Tran} below which there is an explosive onset of spreaders in the population.

Introducing the expression for both contagion rates, $\tilde{\lambda}_A$ and $\tilde{\lambda}_B$, and the total number of individuals infected with each spreading unit, i.e. $\rho_{A_{T}}=\rho_{A}+\rho_{AB}$ and $\rho_{B_{T}}=\rho_{B}+\rho_{AB}$, Eqs.~(\ref{eq:a}-\ref{eq:ab}) turn into~\cite{unidirectional}:
\begin{eqnarray}
\dot \rho_{A_{T}} & = & \rho_{A_{T}}\left[-1+\lambda_Ae^{\gamma(1-\rho_{A_{T}})}(1-\rho_{A_{T}})\right.\nonumber\\
              &  &\left. +\lambda_Ae^{\gamma(1-\rho_{A_{T}})}(\epsilon_{BA}-1)(\rho_{B_{T}}-\rho_{AB})\right]\;,
\label{eq:a_tot}\\
\dot \rho_{B_{T}} & = & \rho_{B_{T}}\left[-1+\lambda_B(1-\rho_{B_{T}})\right.\nonumber \\
             &  & \left. +\lambda_B(\epsilon_{AB}-1)(\rho_{B_{T}}-\rho_{AB})\right]\;,
\label{eq:b_tot}\\
\dot \rho_{AB} & = & -2\rho_{AB}+\epsilon_{AB}\lambda_B (\rho_{A_{T}}-\rho_{AB})\rho_{B_{T}}\nonumber \\
             &  & +\epsilon_{BA}\lambda_Ae^{\gamma(1-\rho_{A_{T}})} (\rho_{B_{T}}-\rho_{AB})\rho_{A_{T}}\;.
\label{eq:ab_tot}
\end{eqnarray}

With this general framework, we can study the effects of interplay between dynamics $A$ and $B$, depending on the coupling interaction rates $\epsilon_{AB}$, $\epsilon_{BA}$, and the explosive nature of dynamics $A$, controlled by the parameter $\gamma$. We start by studying the propagation of two simultaneous infection processes by fixing $\gamma=0$ and setting symmetric interactions, i.e. $\epsilon_{AB}=\epsilon_{BA}=\epsilon$, as sketched in Fig.~\ref{fig:panel1}b. Moreover, we turn the epidemic process $A$ into a social contagion by setting $\gamma \neq 0$ and explore three different couplings summarized in Fig.~\ref{fig:panel1}c: an unidirectional influence of dynamics $A$ into dynamics $B$, a negative feedback between dynamics $A$ and $B$ where dynamics $A$ excites its inhibitor $B$ and a mutual inhibition between both of them.

The exploration of these cases allows us to provide general information on the mechanisms responsible for the emergence of explosive transitions from the interactions between different spreading process. Throughout the manuscript, we characterize such transitions both analytically and with numerical results obtained by integrating the set of equations (\ref{eq:a_tot}-\ref{eq:ab_tot}) until the stationary state is reached. In the numerical simulations, we capture how bistability emerges by performing forward and backward continuations, i.e., by increasing or reducing smoothly the control parameter once stable states are reached and considering as initial condition of the new parameters the final states obtained for the former ones. In case the final states correspond to absorbing configurations, i.e. the spreaders-free population, we add a negligible fraction of infected population to avoid getting trapped there.

\section{Symmetric coupling between two epidemics}
\begin{figure}[t!]
\centering\includegraphics[width=0.7\linewidth]{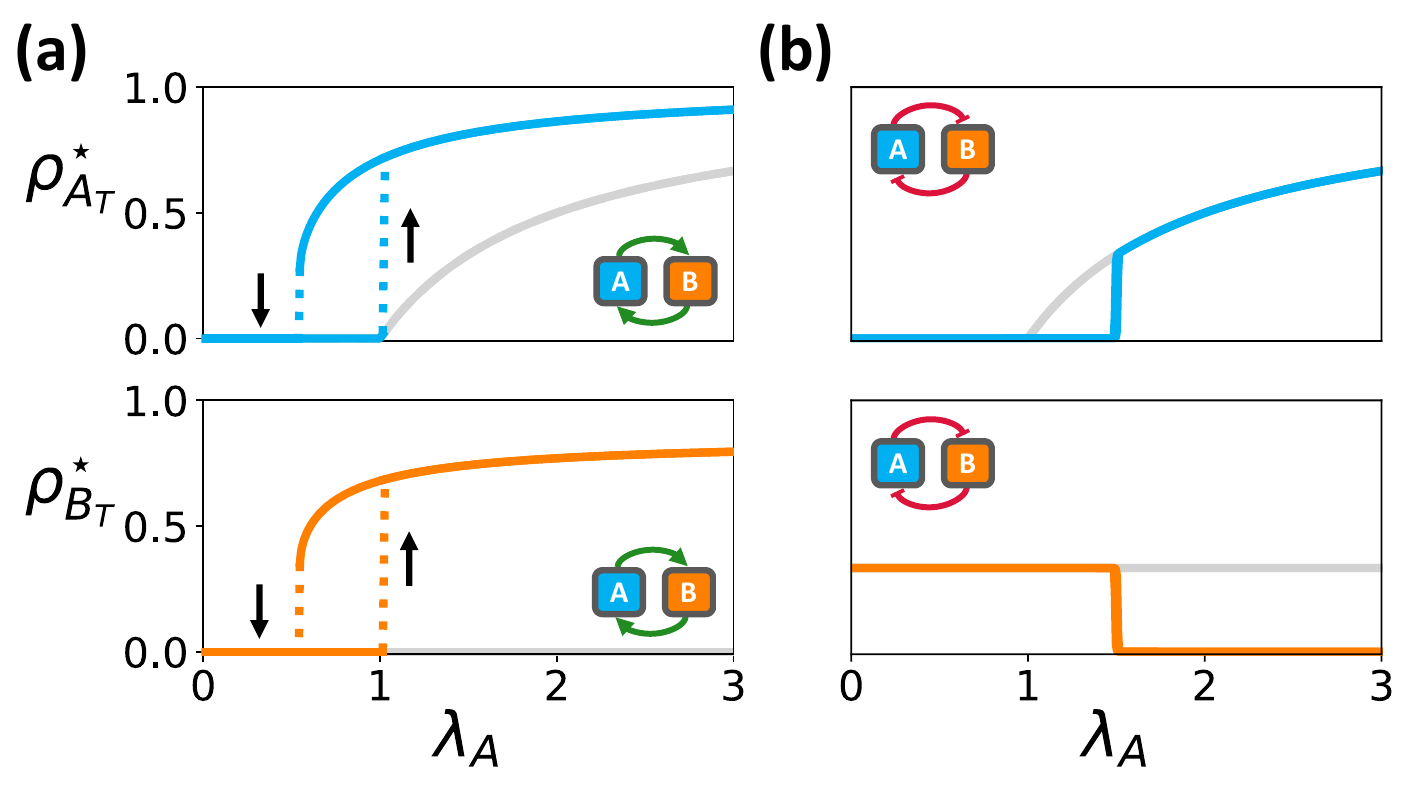}
\caption{\textbf{Epidemic curves for two interacting diseases}. Fraction of population at equilibrium infected by the spreading process $A$, $\rho^*_{A_T}$ (Top row, blue curve), or by the process $B$, $\rho^*_{B_T}$ (Bottom row, orange curve) as a function of the contagion rate $\lambda_A$. The parameters used to simulate the dynamics are $(\lambda_B,\epsilon)=(0.8,8)$ (Panel a) and $(\lambda_B,\epsilon)=(1.5,0)$ (Panel b). In both panels, the grey curves illustrate the expected behavior for two non-interacting spreading processes, i.e. $\epsilon=0$}
\label{fig:panel2} 
\end{figure}

We first focus on two standard SIS dynamics by fixing $\gamma=0$. As stated above, we consider symmetric interactions $\epsilon_{AB}=\epsilon_{BA}=\epsilon$ and explore cooperative ($\epsilon>1$) or mutually exclusive contagion processes ($\epsilon=0$). In Fig.~\ref{fig:panel2}a, we set dynamics $B$ in the absorbing phase by fixing $\lambda_B=0.8$ and explore the effects of increasing $\lambda_A$ in a cooperative regime encoded by $\epsilon=8$. There, we show that mutual cooperation between contagion processes gives rise to the onset of first-order transitions in both epidemic curves. For the sake of comparison, we also represent the curves corresponding to two independent epidemic processes ($\epsilon_{AB}=1$), displaying the typical smooth behavior observed for isolated epidemic dynamics. Moving towards competing dynamics, Fig.~\ref{fig:panel2}b shows that, for $\epsilon=0$, the mutual competition between both spreading processes always leads to the extinction of the least transmissible one. Fixing $\lambda_B=1.5$, we observe that $B$ prevails over $A$ until the latter becomes transmissible enough, i.e $\lambda_A>1.5$, when $B$ becomes extinct. Nonetheless, unlike cooperation, competition does not change the nature of the transitions observed in the dynamics, as pinpointed by the lack of bistability in the epidemic curves.

\begin{figure*}[t!]
\centering\includegraphics[width=1\linewidth]{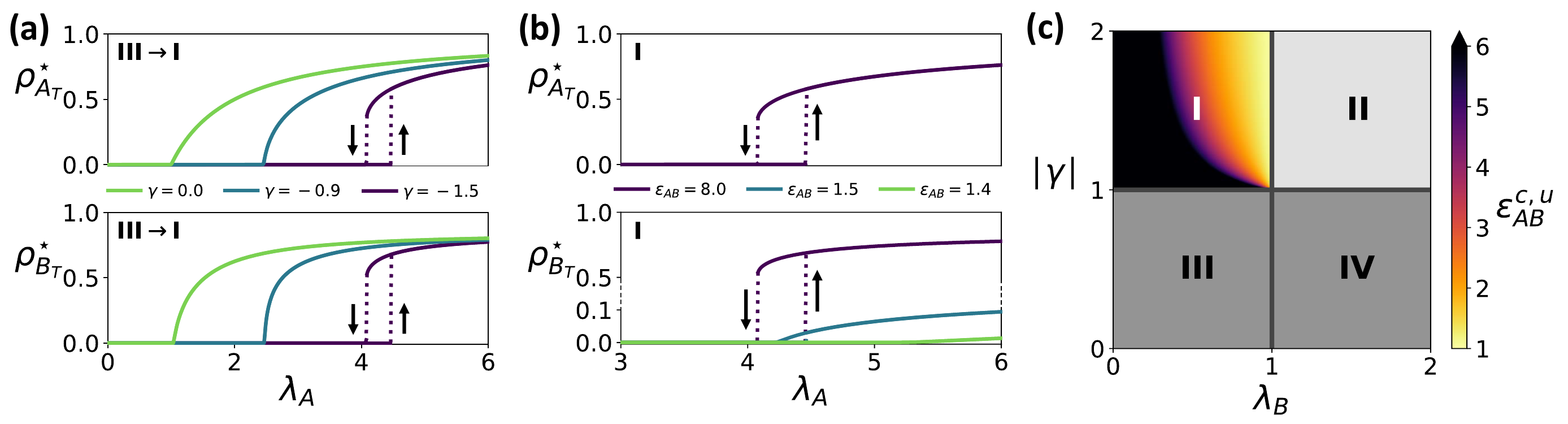}
\caption{\textbf{Pathways to explosive transitions when social dynamics unidirectionally drives epidemic dynamics} (a) Epidemic diagrams $\rho^*_{A_T} (\lambda_A)$ and $\rho^*_{B_T} (\lambda_A)$ for different values of the social pressure $\gamma$ (color code), fixing the coupling strength $\epsilon_{AB}=8$. (b) Epidemic diagrams for different coupling strength $\epsilon_{AB}$ values (color code), fixing $\gamma=-1.5$. In both panels, the instrinsic contagion rate of process $B$ is set to $\lambda_B = 0.8$ and we consider dynamics $A$ to be independent from dynamics $B$, i.e. $\epsilon_{BA}=1$. (c) Phase diagram when coupling unidirectionally social and epidemic dynamics ($\epsilon_{BA}=1$) as a function of the social pressure affecting $A$, $\gamma$, and the contagion rate of dynamics $B$, $\lambda_B$. In region I dynamics $A$ is natively explosive and the activation of $B$ also becomes discontinuous for coupling rates above a given threshold $\epsilon_{AB}^{c,u}$ (color code) given by Eq.~(\ref{eq:eps_crit}). In region II dynamics $A$ is natively explosive and $B$ is in active phase. Thus, for all coupling rates $\epsilon_{AB}$, $B$ undergoes a discontinuous change within its active phase. In regions III and IV, $A$ is no longer explosive and thus the transmitted phenomenology is continuous.}
\label{fig:panel3} 
\end{figure*}

%In Fig.~\ref{fig:panel2}.(a) we study how the coupling rate $\epsilon$ affects the transition underwent by the stationary fractions of infected ($\rho^{\star}_{A_T}$, $\rho^{\star}_{B_T}$) in terms of the infectivities, which are settled equal $\lambda_A=\lambda_B$. As reported previously in the literature \cite{coupledsis}, the cooperation between dynamics yields to a shift on the transition onset, accompanied by a change on its nature: from continuous to discontinuous. In the contrary, when considering in Fig.~\ref{fig:panel2}.(b) full inhibitory interaction, i.e. $\epsilon=0$, only appears the quantitative effect of having a lower fraction of infected in the endemic stationary state.

\section{Coupling social contagion with epidemic processes}
Our previous results retrieve a well-known pathway for explosive transitions: the mutual enhancement of two dynamics presenting second-order transitions in isolation. In this section, we extend our analysis to consider a social dynamics $A$, natively displaying first order transitions, and an epidemic dynamics $B$ characterized by second order transitions. For this purpose, we explore the interplay between the parameter $\gamma$, controlling the nature of the transition observed for dynamics $A$ in isolation, and the interdependencies existing both processes, considering two different scenarios: {\em (i)} dynamics $A$ affects but does not depend on dynamics $B$, referred to as {\em unidirectional coupling}, and {\em (ii)} dynamics $B$ inhibits the transmission of dynamics $A$, denoted by {\em inhibitory coupling}.

\subsection{Unidirectional coupling}
Let us first explore the case in which dynamics $A$ enhances transmissibility of $B$ but remains independent from this dynamical process. Therefore, we now explore the region of the parameters' space delimited by $\epsilon_{AB}>1$ and $\epsilon_{BA}=1$. In Fig.~\ref{fig:panel3}a we show the evolution of the epidemic curves as we vary the parameters tuning the social pressure effect $\gamma$, fixing $\epsilon_{AB}=8$. For $\gamma=0$, we observe how the explosive transition found for mutually cooperative epidemics vanishes when considering unidirectional feedback among them. The effect of the group pressure mechanism, activated by reducing $\gamma$, is two-fold. First, lower $\gamma$ values gives rise to larger epidemic thresholds for the epidemic dynamics $B$, reflecting the latter activation of dynamics $A$ as a result of its hindered transmission by the group pressure effects. More strikingly, when $\gamma$ is small enough, namely $\gamma=-1.5$, such activation takes place abruptly through a first order transition which is transmitted to dynamics $B$. Nevertheless, the transmission of the transition nature hinges on the coupling strength \cite{unidirectional}. To illustrate this phenomenon, we fix $\gamma=-1.5$ and tune $\epsilon_{AB}$, determining how dynamics $A$ enhances dynamics $B$. In Fig.~\ref{fig:panel3}b we show that lower values of the coupling strength ($\epsilon_{AB}=1.4$) give rise to the smooth activation of dynamics $B$ whereas first-order transitions are retrieved when the unidirectional coupling is strong enough.

Fig.~\ref{fig:panel3}c summarizes all posible scenarios found in the parameters space for the outcome of the unidirectional coupling between both dynamical processes. In case dynamics $A$ is not inherently explosive, the transition of dynamics $B$ is continuous, regardless of whether it is inactive (region III) or active (region IV) in isolation. Likewise, in region II, where dynamics $A$ is explosive ($\gamma<-1$) and dynamics $B$ is in the active state ($\lambda_B>1$), there is always a discontinuous transition in $\rho_{B_T}^*$.  In what follows, we focus on region I, where the emergence of first-order transitions in $B$ depends on the interplay between the social pressure and the coupling strength as stated above.

\begin{figure*}[t!]
\centering\includegraphics[width=1\linewidth]{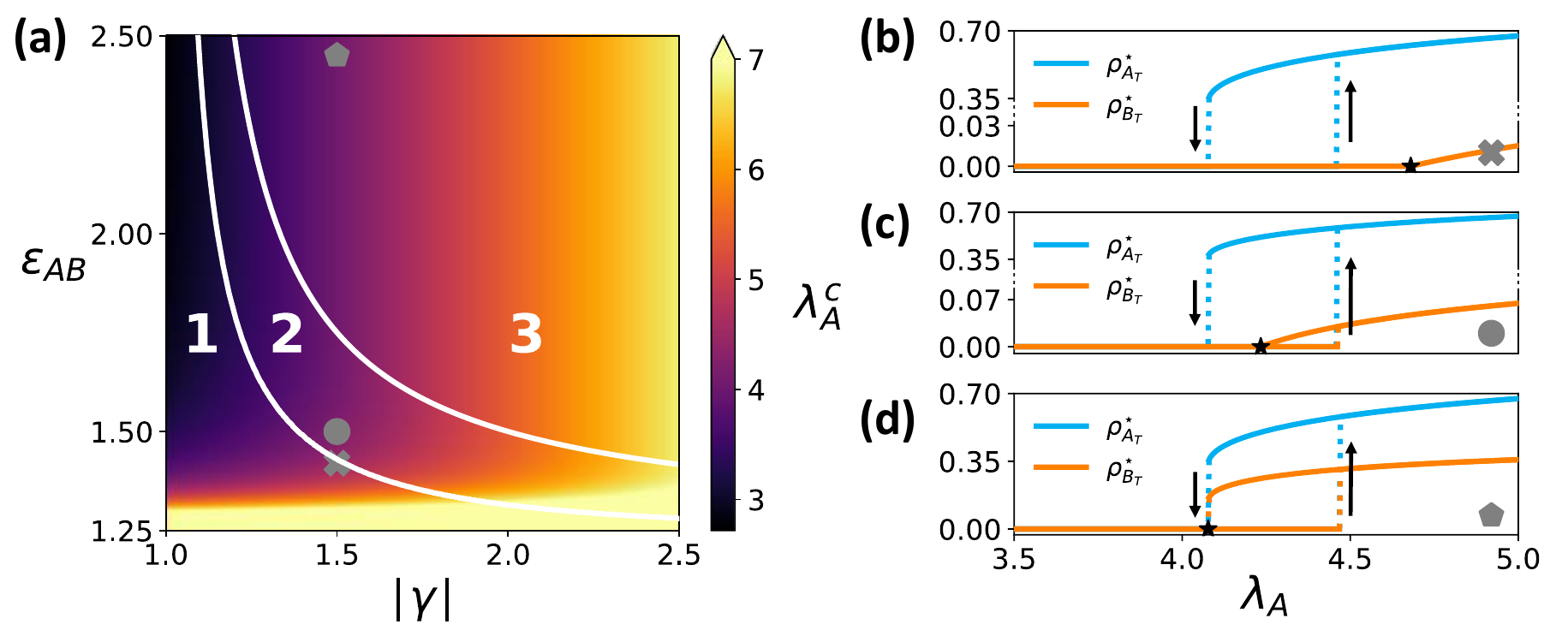}
\caption{\textbf{Onset of epidemics under unidirectional coupling with social dynamics.} (a) Critical values of $A$ infectivity $\lambda_A^c$ to activate dynamic $B$ (color code) as a function of the social pressure  $\gamma$ and the coupling strength $\epsilon_{AB}$ for $\lambda_B=0.8$. This phase diagram presents three different regions corresponding to $\lambda_A^c>\lambda_A^+$ (region 1, continuous transition in $B$), $\lambda_A^-<\lambda_A^c<\lambda_A^+$ (region 2, bistability for $B$ but continuous behavior of the backwards branch), and $\lambda_A^c=\lambda_A^-$ (region 3, bistability and discontinuos jump for dynamics $B$). (b)-(d) Epidemic curves $\rho_{A_T}^*(\lambda_A)$, $\rho_{B_T}^*(\lambda_A)$ for three different points of the parameters space, fixing $(\lambda_B,\gamma)=(0.8,-1.5)$ and considering three different values of the coupling strength to explore the three different regions explained above: $\epsilon_{AB}=1.42$ (Panel b, region 1), $\epsilon_{AB}=1.5$ (panel c, region 2) and  $\epsilon_{AB}=2.5$ (d, region 3). In these panels, the star symbols pinpoints the $\lambda_A^c$ value.}
\label{fig:panel4} 
\end{figure*}

The lack of feedback from dynamics $B$ to dynamics $A$ ($\epsilon_{BA}=1$) enables us to obtain analytical insights into how the transition in the epidemic curves observed at the epidemic threshold depends on both $\epsilon_{AB}$ and $\gamma$. By imposing $\epsilon_{BA}=1$, Eq.~(\ref{eq:a_tot}) becomes
\begin{eqnarray}
\dot \rho_{A_{T}} & = & \rho_{A_{T}}[-1+\lambda_Ae^{\gamma(1-\rho_{A_{T}})}(1-\rho_{A_{T}})]\;,
\label{eq:a_tot_simp}
\end{eqnarray}
which has three fixed points $\rho_{A_T}^\star$ \cite{Exp_Tran} fulfilling $\dot\rho_{A_{T}}(\rho_{A_T}^\star)=0$. The trivial one $\rho_{A_{T}}^{\star,1}=0$ corresponds to the spreaders-free state while the other fixed points are characterized through the use of the Lambert function $W$ as they satisfy:
\begin{equation}
\rho_{A_{T}}^{\star}=1-\frac{1}{\gamma}W\left(\frac{\gamma}{\lambda_A}\right)\;.
\label{eq:solution}
\end{equation}

The Lambert function $W(x)$ is a double-valued function in the interval $x\in(-1/e,0)$, being these values henceforth denoted by $W_0(x)$ and $W_{-1}(x)$ respectively. The range of definition of the Lambert function delimits the lower bound $\lambda_A^-=-\gamma e$ for the emergence of epidemic solutions. Therefore, when $\lambda_A <\lambda_A^-$, the only stable fixed point is the absorbing state $\rho^{*,1}_{A_T}$. Above this threshold, there is a stable solution $\rho_{A_{T}}^{\star,0}$, and another unstable solution $\rho_{A_{T}}^{\star,-1}$ separating the basins of attraction of $\rho_{A_{T}}^{\star,1}=0$ and $\rho_{A_{T}}^{\star,0}$. The first-order transition in dynamics $A$ appears when the former stable state gets positive values at the transition point $\lambda_A^-$. The latter occurs for $\gamma<-1$, as the stable solution at this point reads $\rho_{A_{T}}^{\star,0}(\lambda_A^-)=1+1/\gamma$. Namely, when the social pressure is large enough, a hysteresis cycle spanning the range $(\lambda_A^-,\lambda_A^+)$ appears, where $\lambda_A^+=e^{-\gamma}$ corresponds to the $\lambda_A$ value at which the unstable fixed point $\rho_{A_{T}}^{\star,-1}$ collides with the trivial solution $\rho_{A_{T}}^{\star,1}$ through a saddle node bifurcation, i.e. $\rho_{A_{T}}^{\star,-1}(\lambda_A^+)=0$.

The study of the fixed points of dynamics $A$ allows us to analytically characterize how the social pressure $\gamma$ and the coupling strength $\epsilon_{AB}$ determine the nature of the transition of dynamics $B$ . For this purpose, we impose $\dot\rho_{B_{T}}=0$ in Eq.~(\ref{eq:b_tot}), obtaining the spreaders-free state $\rho_{B_{T}}^{\star,1}=0$ and the implicit relation
\begin{equation}
\rho_{B_{T}}^{\star}=1-\frac{1}{\lambda_B}+\left(\rho_{A_{T}}^{\star}-\rho_{AB}^{\star}\right)\left(\epsilon_{AB}-1\right)\;.
\label{eq:solution_b}
\end{equation}
To study the transition towards an active state for dynamics $B$, let us assume $\rho_{B_{T}}^{\star}=\rho_{AB}^{\star}=0$ and consider the solution $\rho_{A_{T}}^{\star,0}(\lambda_A)$ for the dynamics A.
Plugging the latter conditions into Eq.~(\ref{eq:solution_b}), we obtain:
\begin{equation}
\rho_{A_T}^{*,0} = \frac{1/\lambda_B-1}{\epsilon_{AB}-1}\ ,
\end{equation}
which can be introduced in Eq.~(\ref{eq:a_tot_simp}) to find the value of $\lambda^c_A$ above which $B$ enters in the active phase. The latter reads:
\begin{equation}
\lambda_{A}^c=e^{-\gamma\left[1-\frac{\frac{1}{\lambda_B}-1}{\epsilon_{AB}-1}  \right]}\left[1-\frac{\frac{1}{\lambda_B}-1}{\epsilon_{AB}-1}  \right]^{-1}\;.
\label{eq:lac}
\end{equation}
We represent the former expression in regions 1 and 2 of Fig.~\ref{fig:panel4}a. Each region distinguishes a qualitatively different transition for the dynamics $B$. The nature of the transition depends on the relationship between the $\lambda_{A}^c$ value and the limits of the hysteresis cycles $(\lambda_A^-,\lambda_A^+)$. If $\lambda_A^c>\lambda_A^+$ (region 1), dynamics $B$ presents a second order transition driven by the smooth growth of $\rho_{A_{T}}^{\star,0}(\lambda_A)$ in the supercritical regime, as shown in Fig.~\ref{fig:panel4}b. For $\lambda_A^-<\lambda_A^c<\lambda_A^+$ (region 2), we show in Fig.~\ref{fig:panel4}c that dynamics $B$ exhibits two stable solutions corresponding to the absorbing state $\rho_{B_{T}}^{\star,1}$ and another solution starting from $\rho_{B_{T}}^{\star,0}=0$ at $\lambda_A^c$ and continuously increasing with $\lambda_A$. The limiting case for such behavior, i.e. $\lambda_A^c=\lambda_A^+$, implies that:
\begin{equation}
    1-\frac{\frac{1}{\lambda_B}-1}{\epsilon_{AB}-1}=e^{\gamma\frac{\frac{1}{\lambda_B}-1}{\epsilon_{AB}-1}}\;.
    \label{eq:condi}
\end{equation}

Finally, when $\lambda_A^c=\lambda_A^-$ (region 3), the epidemic curve associated to dynamics $B$ displays a discontinuous transition once dynamics $A$ is activated, as illustrated in Fig.~\ref{fig:panel4}d.  Plugging the latter condition into Eq.~(\ref{eq:lac}), after some algebra we obtain the critical value of the coupling strength $\epsilon^{c,u}_{AB}$ for such discontinous behavior, which reads:
\begin{equation}
\epsilon_{AB}^{c,u}=\frac{\lambda_B+\gamma}{\lambda_B(\gamma+1)}\;.
\label{eq:eps_crit}
\end{equation}
A similar expression for $\epsilon_{AB}^{c,u}$ was obtained in \cite{unidirectional} when modeling the social pressure mechanism as a 3-body simplicial contagion. Denoting the scaled 3-body infectivity by $\lambda_{\Delta}$, the map between both expressions reads $|\gamma|\rightarrow \lambda_{\Delta}^{1/2}$.

\begin{figure*}[t!]
\centering\includegraphics[width=1
\linewidth]{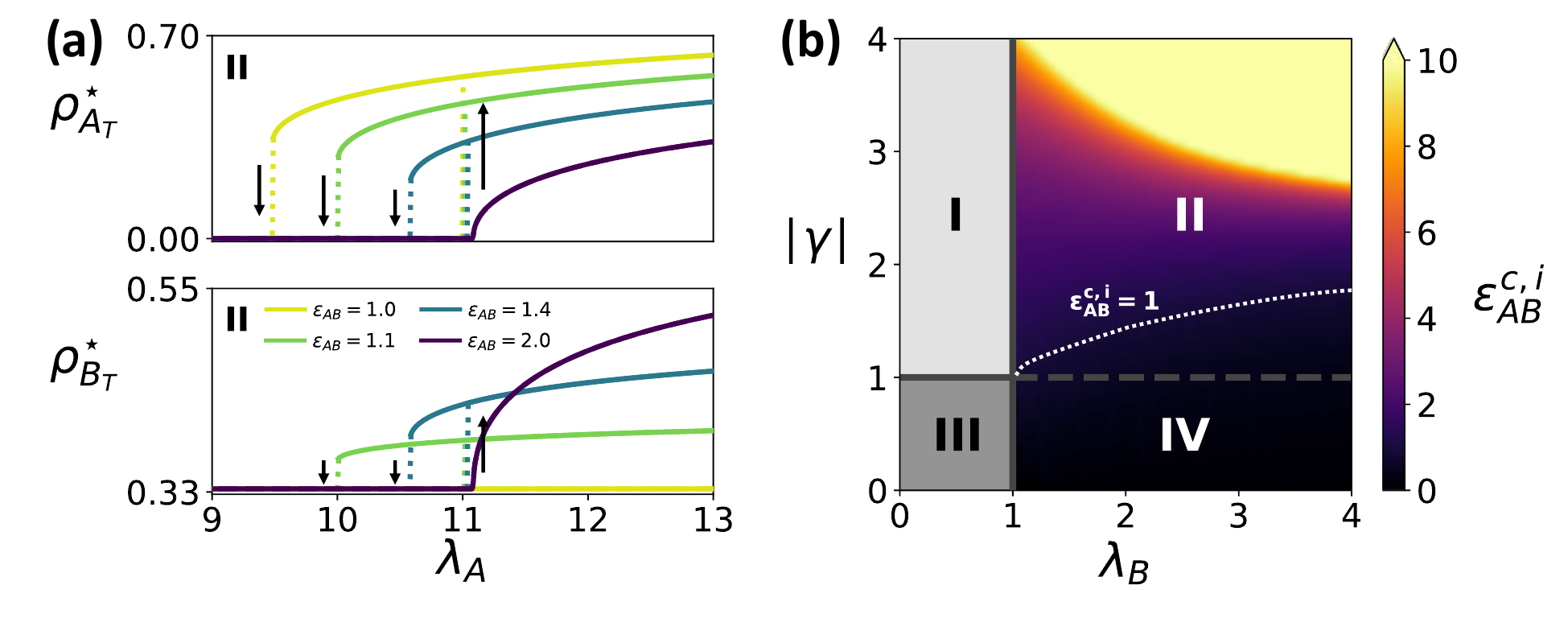}%{Panel_5_new.pdf}%{Fig5.pdf}
\caption{\textbf{Explosive transitions coupling dynamics $A$ with an inhibitor $B$.}
(a) Epidemic curves $\rho_{A_T}^*(\lambda_A)$, $\rho_{B_T}^*(\lambda_A)$ as a function of the coupling strength $\epsilon_{AB}$ when dynamics $B$ inhibits dynamics $A$, i.e. $\epsilon_{BA}=0$, for $\lambda_B=1.5$ and $\gamma=-2$
(b) Phase diagram in presence of inhibitory feedback ($\epsilon_{BA}=0$) in terms of the social pressure for dynamics $A$, $\gamma$, and the infectiousness of dynamics $B$, $\lambda_B$. Regions I and III report a phenomenology alike the unidirectional case: in region I dynamics $A$ is natively explosive and the activation of $B$ becomes explosive above a critical coupling rate, and in region III both $A$ and $B$ dynamics undergo continuous transitions. In regions II, further exemplified in panel (a), and IV, $B$ is in active phase and suppresses the explosive behaviour of $A$ when $\epsilon_{AB}>\epsilon_{AB}^{c,i}$ (color code). Besides, the activation of dynamics $A$ decreases $\rho^*_{B_T}$ for $\epsilon_{AB}<1$ whereas increases the prevalence of both dynamics when $\epsilon_{AB}>1$. $\epsilon_{AB}^{c,i}=1$ is highlighted with a white dotted line. 
}
\label{fig:panel5} 
\end{figure*}

\subsection{Explosive transitions in inhibitory couplings}
So far, we have studied how the social pressure $\gamma$ and the coupling strength $\epsilon_{AB}$ shape the onset of epidemic outbreaks when social contagions are unidirectionally driving epidemic dynamics. In this section, we are interested in extending our analysis to more complex interaction schemes between both dynamical processes. Given their ubiquity in biological and socioeconomic systems, let us consider that both interact through a negative feedback loop. In particular, we assume that dynamics $A$ enhances its inhibitor $B$ by setting $\epsilon_{AB}>1$ and $\epsilon_{BA}=0$, as schematized in Fig.~\ref{fig:panel1}c. In the context of social and epidemic processes, such choice could reflect the interplay between loss of risk perception and contagion dynamics, as the former process leads to stop adopting prevention measures, fostering contagions, which eventually raise awareness in the population~\cite{granell2013dynamical,heinlein2023unraveling}.

\begin{figure*}[t!]
\centering\includegraphics[width=0.9
\linewidth]{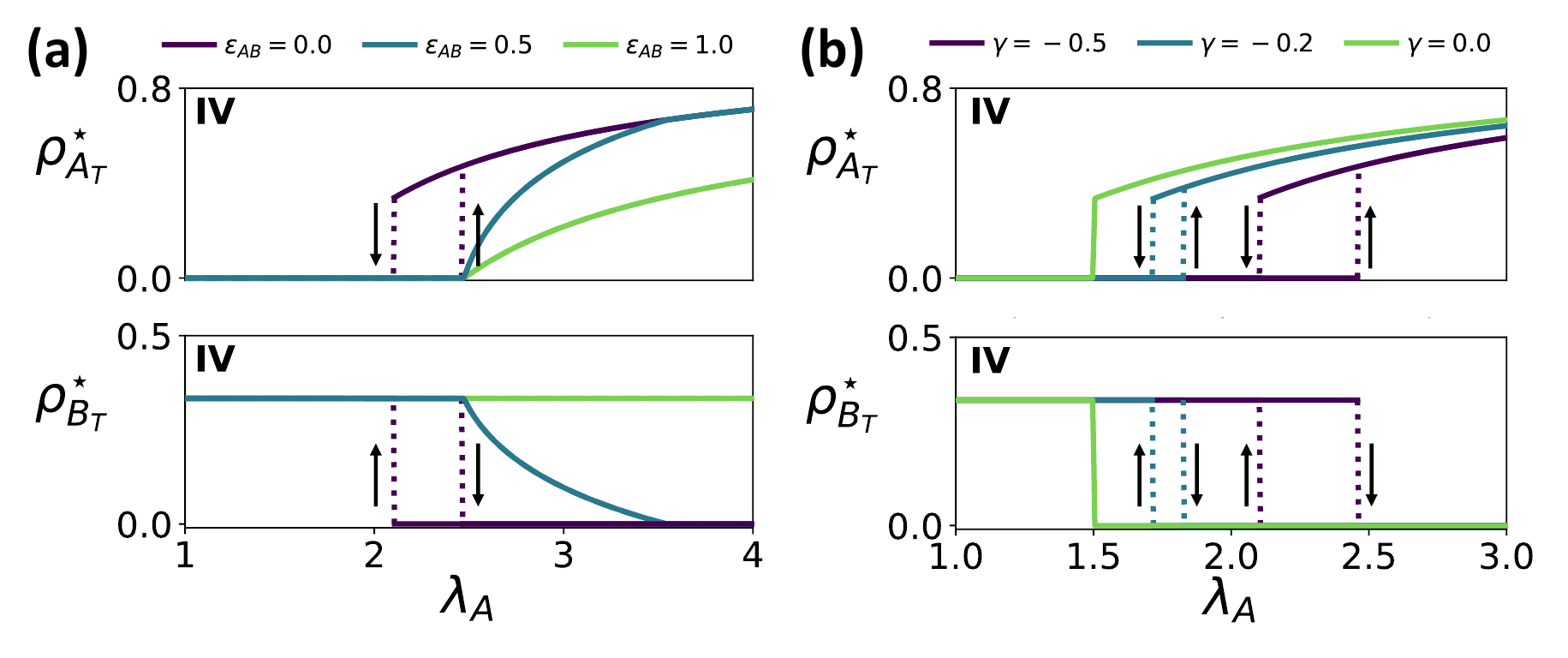}%{Panel_5_new.pdf}
\caption{\textbf{Explosive transitions arising from the mutual inhibition of spreading processes.} (a) Epidemic curves $\rho_{A_T}^*(\lambda_A)$, $\rho_{B_T}^*(\lambda_A)$ as a function of the coupling strength $\epsilon_{AB}$ fixing $\gamma=-0.5$. (b) Epidemic curves $\rho_{A_T}^*(\lambda_A)$, $\rho_{B_T}^*(\lambda_A)$ as a function of the social pressure $\gamma$ fixing $\epsilon_{AB}=0$. In both cases, we set $(\epsilon_{BA},\lambda_B)=(0,1.5)$}.
%\santiago{1) dos figuras; una con a y c, y la otra con b. 2) fig c con solo epsilons mayores que 1. Y otra curva más en la b DEJO EL BOCETO PARA TENER LA IDEA}
\label{fig:panel6} 
\end{figure*}

We represent in Fig.~\ref{fig:panel5}a the epidemic curves for different coupling strength $\epsilon_{AB}$, fixing $\lambda_B=1.5$ and $\gamma=-2$, thus considering that $B$ is in the active state and $A$ presents a first-order transition in isolation. In the case of $B$ being independent from $A$, $\epsilon_{AB}=1$, the prevalence of $B$ remains constant while $A$ preserves the first-order transition between the active and inactive states. Remarkably, the presence of the inhibitor $B$ hampers the onset of such transition by reducing the pool of susceptible individuals prone to contract $A$. In particular, the threshold destabilizing the absorbing state of dynamics $A$ reads $\lambda_A^{C,i}=e^{-\gamma}/(1-\rho^*_B)$. As the coupling is activated by increasing $\epsilon_{AB}$, we observe a discontinuous transition for dynamics $B$ with the presence of hysteresis cycles and bistable solutions. Nevertheless, such cycles start shrinking as the negative feedback loop structure gains relevance through the increase of the coupling strength $\epsilon_{AB}$. Eventually, large values of $\epsilon_{AB}$ prompt second order transitions in both $A$ and $B$, thus removing the inherent explosive nature of the transitions observed for dynamics $A$ in isolation. This result highlights the relevance of negative feedback loops to ensure systems robustness against external perturbations.
 
Analogously to the unidirectional case, we now provide a general description of the transitions observed in both dynamics in the space ($\lambda_B$,$\gamma$) in Fig.~\ref{fig:panel5}b. We observe four different regions: First, in regions I and III, dynamics $B$ is not able to reach the endemic state by itself, as $\lambda_B <1$. Thus, when $B$ is activated by $A$, the qualitative behaviour of the driver dynamic $A$ is not affected, i.e. in region I (III) $A$ undergoes a discontinuous (continuous) transition. Region II captures the phenomenology described for Fig.~\ref{fig:panel5}a where the existence of negative feedback loops between both dynamics might change the nature of their transitions. To further characterize such phenomenon, we numerically obtain the value of the critical coupling strength $\epsilon^{c,i}_{AB}$ above which smooth transitions appear in the epidemic curves. Such value increases with the infectivity $\lambda_B$, enhancing the presence of the inhibitor $B$ in isolation, and the social pressure $|\gamma|$, increasing the abruptness of the intrinsic transitions for dynamics $A$. Conversely, decreasing $|\gamma|$ reduces the latter abruptness, thus diminishing the role of negative feedbacks in ensuring the smooth behavior of the epidemic curves. In particular, in Region IV, characterized by an active $B$ dynamics ($\lambda_B>1$) and a non-explosive dynamics $A$ ($\gamma>-1$), such smooth transitions are also observed in mutually competitive scenarios, as $\epsilon^{c,i}_{AB}<1$.

To further characterize the Region IV, we represent in Fig.~\ref{fig:panel6}a, the evolution of the epidemic curves for different values of the coupling strength $\epsilon_{AB}$ when fixing $\gamma=-0.5$. There we observe that in presence of weak competition, $0<\epsilon_{AB}<1$, dynamics $A$ smoothly activates at $\lambda_A^{C,i}=e^{-\gamma}/(1-\rho^*_B)$ giving rise to a smooth decrease of the prevalence of dynamics $B$. As competition is strengthened, such behavior becomes more abrupt until giving rise to a first-order transition between the existence and extinction of dynamics $B$ in the case of mutually exclusive processes ($\epsilon_{AB}=0$). Note that hysteresis cycles are driven by the social pressure effect $\gamma$, as this behavior is not present when considering competing epidemics as proven in Fig.~\ref{fig:panel2}b. The relevance of the synergistic contagion rate is further illustrated in Fig.~\ref{fig:panel6}b, where we show that decreasing the social pressure for exclusive competing processes ($\epsilon_{AB}=\epsilon_{BA}=0$) shrinks the bistability area, which eventually vanishes in absence of this mechanism, i.e $\gamma=0$.
\section{Conclusions}

Understanding the emergence of explosive phenomena is of paramount importance, since the abrupt nature of those processes amplifies the impact of perturbations on their equilibria~\cite{lamata2023collapse}. In this study, we have used a minimal compartmental model to explore different pathways for the emergence of explosive transitions in the context of two interacting spreading dynamics. Our model couples an epidemic dynamics with a socially-inspired spreading process with a synergistic contagion rate modeling the impact of social pressure on transmission. Our framework allows exploring a myriad of scenarios for intertwined dynamics by tuning both their individual properties, i.e. their dynamical states and transitions in isolation, and their interdependencies. 

In absence of any social pressure, our model retrieves well-known results studied in the context of interacting epidemics. Namely, discontinuous explosive transitions might arise from the mutual excitation of the dynamics whereas the mutual exclusion of competing pathogens leads to a single equilibria characterized by the extinction of the least transmissible process. Introducing the social pressure mechanism through the synergistic contagion rate gives rise to a much richer phenomenology. First, we explore under which conditions first-order transitions can be observed for epidemic dynamics unidirectionally driven by an underlying social process. We reveal that the social dynamics must present a discontinuous transition in isolation to observe abrupt ones in the coupled system. In these conditions, such discontinuous transitions always emerge for active epidemics whereas an interplay between the social pressure and the coupling strength determines the presence of abrupt transitions for inactive outbreaks in isolation.

As discussed above, first-order transitions represent a hazard for systems' stability under external perturbations. For this reason, we have also explored alternative interaction schemes to turn first-order transitions observed for the isolated dynamics into smooth ones in the coupled system. Our numerical results reveal that such outcome can be achieved through negative feedback loops where the dynamics presenting first-order transitions enhances its dynamical inhibitor, which displays smooth behavior and should be in the active phase in isolation. In particular, the coupling strength needed to switch off first-order transitions always increases with the abruptness of such transitions and shows a non-trivial dependence with the dynamical state of the inhibitor. This finding supports the major role of negative feedback loops in guaranteeing the robustness and stability of biological systems~\cite{ferrell2013feedback,alon2007network}. Indeed, our findings reproduce the impact of coupling negative and positive feedback loops in gene regulatory networks~\cite{tian2009interlinking,ferrell2002self}. Finally, we also find that synergistic contagion rates give rise to explosive transitions for mutually competing spreading pathogens, mimicking the results observed for competing dynamics on higher-order networks~\cite{LI2022126595}.

Summing up, this work provides a general atlas for the emergence of explosive transitions in intertwined spreading processes. For the sake of analytical tractability, our model considers well-mixed populations, yet exploring the interplay between the network of dynamical interdependencies and the complex structure of contacts driving spreading processes remains future work. Moreover, we believe it is also worth extending this analysis to other dynamical processes, such as synchronization~\cite{arola2022emergence,soriano2019explosive} or social dilemmas~\cite{szolnoki2009resolving}, to further explore the physics arising from interacting dynamics. Overall, our results underscore the crucial role of interdependencies in changing the nature of the transitions and the dynamical states observed for contagion processes in isolation, thus calling for their inclusion when modeling their simultaneous propagation.

\section*{Acknowledgements}
S.L.O and J.G.G. acknowledge financial support from MICIN through project PID2020-113582GB-I00/AEI/10.13039/501100011033, and from the Departamento de Industria e Innovaci\'on del Gobierno de Arag\'on y Fondo Social Europeo (FENOL group grant E36-23R). D.S.P acknowledges the financial support of the Calouste Gulbenkian Foundation through the PONTE program.\\

\bibliography{apssamp}
\end{document}